# Memristive Charge-Flux Interaction Still Makes It Possible To Find An Ideal Memristor

**Frank Z. Wang**

Division of Computing, Engineering & Mathematics, University of Kent, UK (frankwang@ieee.org)

**ABSTRACT** In 1971, Chua defined an ideal memristor that links charge $q$ and flux $\varphi$. In this work, we demonstrated that the direct interaction between physical charge $q$ and physical flux $\varphi$ is memristive by nature in terms of a time-invariant $\varphi$-$q$ curve being nonlinear, continuously differentiable and strictly monotonically increasing. Nevertheless, this structure still suffers from two serious limitations: 1, a parasitic "inductor" effect, and 2. bistability and dynamic sweep of a continuous resistance range. Then we discussed how to make a fully-functioning ideal memristor with multiple or an infinite number of stable states and no parasitic inductance, and gave a number of suggestions, such as "open" structure, nanoscale size, magnetic materials with cubic anisotropy (or even isotropy), and sequential switching of the magnetic domains. At last, we responded to a recent challenge from arXiv.org that claims that the structure reported in our retracted JAP paper "is simply an inductor with memory" since it did not pass their designed capacitor-memristor circuit test. Contrary to their conjecture that "an ideal memristor may not exist or may be a purely mathematical concept", we remain optimistic that researchers will discover an ideal memristor in nature or make one in the laboratory.

**INDEX TERMS** circuit theory; circuit element; ideal memristor; neuromorphic computing; brain-inspired computer.

## I. INTRODUCTION

Chua defined the ideal memristor in 1971, which directly interacts physical flux $\varphi$ and physical charge $q$ [1], analogous to the resistor, which directly interacts physical voltage and physical current; the capacitor, which directly interacts physical voltage and physical charge; and the inductor, which directly interacts physical current and physical flux.

From a rigorous physics-theoretical perspective, the HP memristor [2] is too incomplete (no magnetic flux), too complex (a "sandwich" structure), and too special (even a chemical reaction in the memory-holding oxygen vacancies) to be fundamental.

The HP memristor lacks a magnetic flux term in the original memristor definition. Williams from HP thought the actual definition of memristance would be more general. They argued, "Linking electric charge and magnetic flux is one way to satisfy the definition, but it's not the only one. In fact, it turns out you can bypass magnetic interaction altogether." [3].

Can we truly bypass magnetic interaction to define a memristor?

Unfortunately, thus far, it was somewhat popular to define "$\varphi$" and "$q$" in a purely mathematical way without giving "$\varphi$" and "$q$" any physical interpretations: "$\varphi$" was defined as the time integral of voltage "$v$" and "$q$" was defined as the time integral of current "$i$". Most people simply "abused" Chua's suggested fingerprint [4] "If it (the $v$-$i$ hysteresis loop) is pinched, it is a memristor" by ignoring its prefixing (from an experimental perspective) and its context ("This definition greatly broadens the scope of memristive devices … into three classes: Ideal Memristors, Generic Memristors, and Extended Memristors.") [4].

At least three examples can be given to demonstrate that it is highly risky to define a memristor in the $v$-$i$ plane.

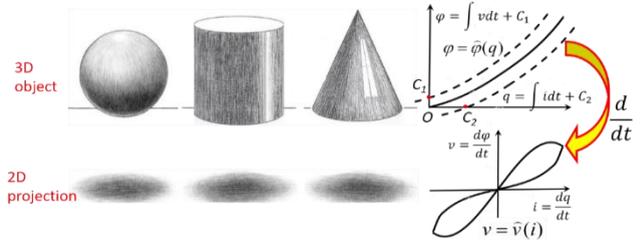

**Fig.1 The conformal differential transformation is similar to the 3D object projection in terms of projecting something from a "high-dimensionality" space into a "low-dimensionality" space.**

The first example is the indefinite integration itself. Starting from $v$, we have $\int v(t)dt = \varphi(t) + C$, where $C$ is an arbitrary constant [meaning that any value for $C$ makes $\varphi(t) + C$ a valid antiderivative of $v(t)$]. This constant expresses an ambiguity inherent in the construction of integration; there is an indefinite number of antiderivatives of $v(t)$, and therefore, a given $v$-$i$ curve (as postulated by Chua [1], it's a pinched hysteresis loop (looking like a diagonal "∞") cannot uniquely determine the $\varphi$-$q$ curve, as shown in the inset of Fig.1.

In other words, the $\varphi$-$q$ plane is a space with "high-dimensionality" (if the charge-flux relationship of a memristor is a high-order polynomial as imagined by Chua [5]), whereas the $v$-$i$ plane is its "dimensionality-reduced" space. The conformal differential transformation [4][5] projecting the $\varphi$-$q$ curve into the $v$-$i$ plane just plays the role of dimensionality reduction (e.g., $\frac{dx^n}{dx} = nx^{n-1}$). Imagine that we are shining a light from above a 3D object and looking at the shadow it casts on a 2D screen. The transformation from 3D to 2D is unique, but the opposite is not; we cannot work out whether the original 3D object is a ball, cylinder or cone when our viewed 2D projection is a circle. A schematic (Fig.1) best illustrates this. We cannot simply restore the lost information in the projection in a single "anti-projection".

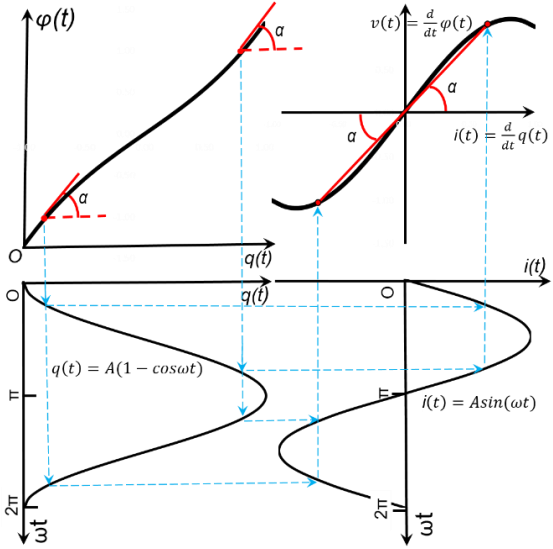

**Fig. 2 No hysteresis can be seen in the *v-i* curve for an ideal memristor with an odd-symmetric *φ-q* curve.**

The second example is that, as shown in Fig.2, no hysteresis can be seen in the *v-i* curve if the *φ-q* curve is odd-symmetric, although it complies with all the three criteria [1] for the ideal memristor's *φ-q* curve: 1. nonlinear, 2. continuously differentiable, and 3. monotonically increasing. The *φ-q* curve projects to the corresponding *v-i* curve via the so-called conformal differential transformation (for a periodic input current): obtain the angle of incline $\alpha$ of the tangent line at an operating point in the $\varphi = \hat{\varphi}(q)$ curve and draw a straight line through the origin in the *v-i* plane whose angle of incline is also $\alpha$.

The third example is that, as shown in Fig.3, a pinched *v-i* curve that is asymmetric even results from a double-valued *φ-q* curve, which indicates that the corresponding memristor is not ideal at all. It is concluded that an ideal memristor that is originally defined in the constitutive *φ-q* plane should not be characterized in the *v-i* plane.

What we have observed from the above three examples is that "even if it's pinched it may not be an ideal memristor". In our opinion, it is bad to define the ideal memristor in the *v-i* plane, it is worse to have no physical magnetic flux, and it is even worse to pretend to have a magnetic flux that was virtual and calculated from other physical attributes. Strictly speaking, it is incorrect to bypass or replace the direct, physical charge-flux interaction with a "virtual" interaction between the charge (that is normally physical, e.g., oxygen vacancies as ionic current in the HP memristor [2]) and the integral over the voltage (no choice due to the lack of magnetic flux).

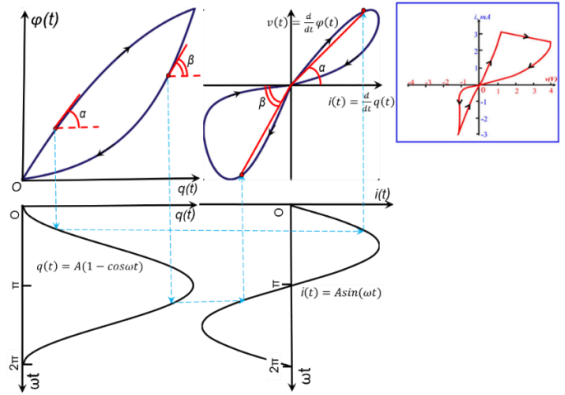

**Fig.3 A nonideal memristor even has a pinched *v-i* hysteresis loop.** This memristor has two *φ-q* characteristic branches, each of which is chosen depending on the polarity of the input current. Some practical devices exhibit such *q–φ* curves [6].

Other skeptics also expressed a similar concern regarding the lack of a charge-flux interaction in the HP memristor. As mentioned in the same issue of IEEE Spectrum, as early as in 2009, skeptics argued that the HP memristor is not a fourth fundamental circuit element but an example of bad science [7]. In 2015, Vongehr even declared, "The Missing Memristor Has Not Been Found" (the title of their Nature Scientific Reports paper) in the sense that an ideal memristor device should be grounded in fundamental symmetries of basic physics, here electromagnetizm and that the "ideal/real/perfect/… memristor" needs magnetizm [8]. It is worth mentioning that the Vongehr work [8] is only one facet of the opposition to the ideality of a new passive fundamental electrical component. There are other reputed and more recent studies [9][10][11].

## II. A STRUCTURE WITH CHARGE-FLUX INTERACTION
In order to design an artificial device with a direct interaction between physical charge and physical flux and then prove that the interaction is memristive, first principles originated by Aristotelians more than 2300 years ago [12] may help. That is to say, we should start directly at the level of the established memristor concept based on physical charge and physical flux without making any assumption such as an empirical model (e.g., a pinched hysteresis voltage-current loop) and parameter fitting (e.g., voltage as a derivative of flux, current as a derivative of the charge).

After numerous failures, the structure in Fig.4 was eventually designed to introduce the direct charge-flux interaction. Through the conductor that carries a current and the magnetic lump that hosts a magnetization, the amount of the charge can be controlled by the time interval of the current flow. The flux can be adjusted by the magnetization rotation. In order to pick up a signal from the charge-flux coupling, this conductor can simultaneously play two roles; on the one hand, it carries a current to switch the magnetization in the lump, and on the other hand, it can also sense the possibly induced voltage by the switched flux.

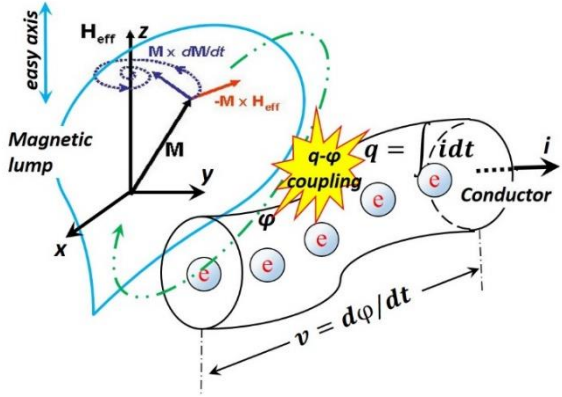

**Fig.4 The charge-flux interaction in a structure with a magnetic lump and a current-carrying conductor.** The Oersted field generated by the current *i* rotates or switches the magnetization *M* inside the lump and consequently the switched flux *φ* induces a voltage *v* across the conductor, resulting in a changed (equivalent) memristance. The LLG model of flux reversal is also shown. If the magnetic field $H_{eff}$ is applied in the Z direction, the saturation magnetization vector M(t) follows a precession trajectory from its initial position ($m_0$≈-1) until ($m$≈1), i.e., the magnetization *M(t)* reverses itself and is eventually aligned with the magnetic field $H_{eff}$.

### III. LLG MODEL OF FLUX REVERSAL
The resulting equivalent resistance, the ratio between the sensed voltage and the driving current, of the above structure is possibly memristive. It would be highly unlikely to have a linear charge-flux interaction due to the existence of magnetic material with rich hysteresis. That is, it is the magnetic lump material that may provide a source of nonlinearity necessary for such a structure.

In order to describe the charge-flux interaction in a mathematically and physically rigorous way, the Landau–Lifshitz–Gilbert equation [13][14] was used. Assuming $m(t) = \frac{M_Z(t)}{M_S}$, in which $M_Z$ is the component of the saturation magnetization $M_S$ in the Z axis, and a magnetic field *H* is applied along *Z*. The model is expressed as below:
$$m(t) = tanh\left[\frac{q(t)}{S_W} + C\right], \quad (1)$$
in which $S_W$ is a switching coefficient and *C* is a constant of integration such that $C = tanh^{-1}m_0$ ($m_0$ is the initial value of *m*) if *q(t=0)=0* (no accumulation of charge at any point).

By Faraday's law, the induced voltage *v(t)* is:
$$\mu_0 S \frac{dM_Z}{dt} = S \frac{dB_Z}{dt} = \frac{d\varphi_Z}{dt} = -v(t), \quad (2)$$
where $\mu_0$ is the permeability of free space and *S* is the cross-sectional area.

From Eq.2, we obtain
$$\varphi = \mu_0 SM + C' = \mu_0 S M_S m + C', \quad (3)$$
where $C'$ is another constant of integration.

Assuming $\varphi(t=0) = 0$, we have $C' = -\mu_0 S M_S m_0$, so
$$\boxed{\varphi = \mu_0 S M_s \left[tanh\left(\frac{q}{S_W} + tanh^{-1}m_0\right) - m_0\right] \triangleq \hat{\varphi}(q). \quad (4)}$$

Eq.4 complies with the new three criteria [6] for the ideal memristor: 1. Nonlinear; 2. Continuously differentiable; 3. Strictly monotonically increasing. Fig.5 shows a typical *φ-q* curve with $m_0$=-0.964 (such a value reflects the intrinsic fluctuation otherwise *M* will stick to the stable equilibriums $m_0 = \pm 1$).

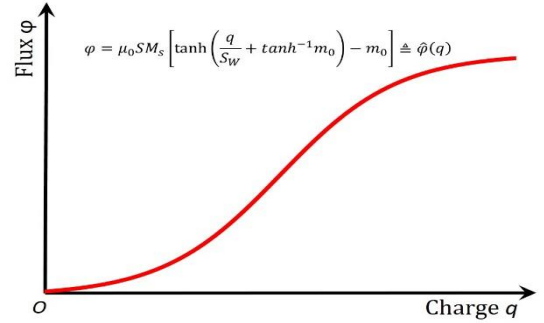

**Fig.5 Intuitively, the S-shaped φ-*q* curve (Eq.4) of this structure vividly depicts the self-limiting charge-flux interaction in a circuit element.** It complies with the three criteria for the ideality of a memristor [1][6]: a. nonlinear; b. continuously differentiable; c. strictly monotonically increasing.

Fig.5 agrees with those experimentally observed *φ-q* curves in the magnetic cores [15][16][17]. Therefore Eq.4 is used as a constitutive curve in this work. The constitutive curve in Fig.4 is dramatically different from that of Chua's "fictitious" memristor with a charge-flux relationship $\varphi = \frac{1}{3}q^3 + q$ in his tutorial [5]. We have $\lim_{q\to\pm\infty} tanh(q) = \pm 1$ for a hyperbolic function whose output range is normalized from -1 to 1 (no matter how big the input is) whereas $\lim_{q\to\pm\infty}\left(\frac{1}{3}q^3 + q\right) = \pm\infty$ for a polynomial function (that diverges to infinity). In other words, such a structure's operation range is finite (where $M(q) = \frac{d\varphi}{dq} \neq 0$) whereas the "fictitious" memristor's range is infinite. Such a self-limiting hyperbolic tangent function is more natural than other functions since the *S*-curve approaches 1 as *x* is +∞ and approaches zero as *x* is −∞. In biology and ecology, a self-limiting colony of organisms limits its own growth by its actions (releasing waste that is toxic to the colony once it exceeds a certain population) [18]. In this instance, there is a clear physical explanation for the saturation of this structure: it is because the magnetization vector is as aligned as the magnetic field allows it to be. The change in the magnetization alignment is negligible on increasing the field above this.

### IV. PARASITIC INDUCTANCE & STEPWISE MEMRISTANCE
Inevitably, a parasitic inductance of this structure in Fig.4 coexists, especially if it is a core with a high coupling efficiency (Fig.6), which can be described by:
$$L = 0.4\pi\mu N^2 \frac{A}{l} \times 10^{-5} (mH), \quad (5)$$
where *L* is the inductance (*mH*) of the magnetic core, $A = \frac{D-d}{2} \cdot h \ (cm^2)$ is the cross-sectional area of the core, $l = \pi \cdot \frac{D+d}{2} \ (cm)$ is the average length of the core, *μ* is the permeability of the core, *N* is the number of the turns in the coil (*N=1* in Fig.4), *D (cm)* is the outer diameter of the core, *d (cm)* is the inner diameter of the core, and *h (cm)* is the height of the core [19].

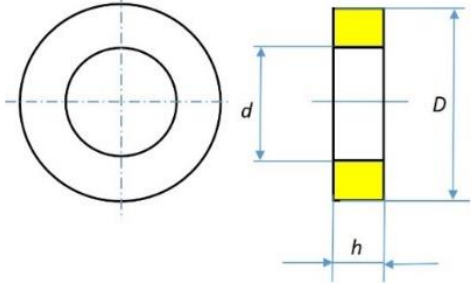

**Fig.6 Geometry of a magnetic core for the inductance calculation.**

If a fixed aspect ratio *D=2d=2h* is taken, Eq.5 can be simplified to:
$$L = 0.4\pi\mu N^2 \frac{h}{3} \times 10^{-5} (mH), \quad (6)$$
which implies that the inductance of a magnetic core scales with its physical size. This is truly encouraging in the sense that, in principle, an ideal memristor on the nanoscale is expected to have a negligible parasitic inductance.

Notably, as shown in Eq.6, the parasitic inductance in this device is not a function of the charge (otherwise it will become a mem-inductor [20]), whereas its resistance (memristance) is a function of the charge.

An antiferromagnet/ferromagnet heterostructure [21] was found to exhibit a stepwise memristance (due to the sequential switching of the domains in the ferromagnetic layer), in which one can freeze the resistance statically at any intermediate time point. It exhibits a continuous "state-dependent Ohm's law" and provides a solution of Chua's Enigma: all non-volatile memristors have continuum memories [22].

### V. CONCLUSIONS & ARGUMENTS
Our work represents a step forward in terms of verifying the memristive charge-flux interaction but we have not reached the final. The structure has two serious limitations:
1. The aforementioned memristivity fingerprint hides behind a superficial inductor effect due to its inductor-like structure. It was necessary to apply a constant input current (such as a step-function or a sequence of square-wave pulses) to depress the inductor effect ($\because v = L\frac{di}{dt} = 0$). For this reason, the structure reported in Ref.[26][27] failed to pass the capacitor-memristor circuit test (as a charge-tracking device), in which a sawtooth, a sinusoidal or another "nonflat" periodic input current is applied [23][24][25]. Nevertheless, it is unfair to conclude that the prototype [26][27] is "simply an inductor with memory" [23]. Despite the existence of a parasitic inductance, the structure displays memristivity; similarly, a real-world resistor is still thought to be a resistor despite the existence of an (inevitable) parasitic inductance and/or capacitance. Most importantly, the structure exhibits that its charge-flux interaction is memristive by nature.

2. The structure is bistable and dynamically sweeps a continuous range of resistances [26][27]. This "dynamical continuity" results from the uniaxial magnetic anisotropy of the prototype, which contains magnetic material with only one easy axis. A fully-functioning ideal memristor should have multiple or an infinite number of stable states so its static memristance can be "frozen" at any intermediate point in time.

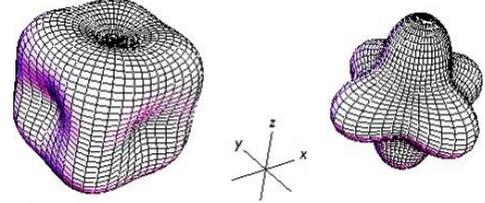

**Fig.7 Cubic anisotropy energy surfaces.** The magnetic moment in cobalt has a dependence of energy level towards one particular direction (the easy axis) then it has uniaxial anisotropy. The biases in iron (left) and nickel (right) are towards many particular directions, then they have multiple easy axes and possess cubic anisotropy [28].

Limitation 2 ("dynamic continuity") may be overcome by using magnetic materials with cubic anisotropy (three or four easy axes as shown in Fig.7) or even a magnetically isotropic material (no preferential direction for its magnetisation) [28]. In addition, an antiferromagnet/ferromagnet heterostructure exhibited stepwise memristance due to sequential switching of the domains in its ferromagnetic layer [21]. However, this bilayer is nonideal since its resistance is a function of several internal state variables, including temperature, and it is not a function of only charge [21].

It is worth mentioning that here we did not use the so-called rotational model [14][15][16] implemented in Ref. [26], which was unfortunately retracted because, according to the JAP Editors, "This model determines only the equilibrium states of *m* and not the time evolution of the switching at the critical fields. The switching process can be indeed described by the LLG equation." [27]. Notably, the JAP Editors did not challenge our experiments, which showed that our device reported in [26] exhibited a number of fingerprints for a memristor. The above formulas in this submission were directly derived from the LLG equation as suggested by the JAP editors [27].

A fully-functioning charge-flux-interaction-based ideal memristor with multiple or an infinite number of stable states and no parasitic inductance is still highly in demand in terms of filling the gap of 50 years [1][7][8]. The existence of such a fundamental circuit element may appeal to many researchers in the memristor field within the context of the theoretical circuit innovations that depend on charge-flux linkage [29][30][31][32]. We are still optimistic that researchers will discover an ideal memristor in nature or make one in the laboratory, although some researchers felt that an ideal memristor may not exist or may be a purely mathematical concept [25].